\documentclass[prb,twocolumn,amsmath,amssymb,floatfix,showpacs,superscriptaddress]{revtex4}

\usepackage{graphicx}
\usepackage{dcolumn}
\usepackage{times}
\usepackage{bm}

\begin{document}

\preprint{APS/123-QED}

\title{Phonon Spectroscopy  by Electric Measurements of Coupled Quantum Dots}

 \author{A. Ueda}
 \email{akiko@bgu.ac.il}

 \affiliation{Department of Physics, Ben Gurion University,
Beer Sheva 84105, Israel}
 \author{O. Entin-Wohlman}

\altaffiliation{Also at Tel Aviv University, Tel Aviv 69978,
Israel}

\affiliation{Department of Physics, Ben Gurion University, Beer
Sheva 84105, Israel}

\author{ M. Eto}

\affiliation{Faculty of Science and Technology, Keio University,
3-14-1 Hiyoshi, Kohoku-ku, Yokohama 223-8522, Japan
}

\author{A. Aharony}

\altaffiliation{Also at Tel Aviv University, Tel Aviv 69978,
Israel}
\affiliation{Department of Physics, Ben Gurion University, Beer
Sheva 84105, Israel}

\date{\today}
\begin{abstract}
We propose phonon spectroscopy by electric measurements
of the low-temperature conductance of coupled-quantum dots,
specifically employing dephasing of the
quantum electronic transport by the phonons.
The setup we consider consists of
a T-shaped double-quantum-dot (DQD)
system in which only one of the dots (dot 1)
is connected to external leads and the other (dot 2) is coupled solely 
to the first one.  For noninteracting
electrons, the differential conductance of such a system vanishes
at a voltage located in-between the energies of the bonding and
the anti-bonding states, due to destructive interference. When
electron-phonon (e-ph) on the DQD is invoked, we find that, at low temperatures,
phonon emission taking place on dot 1 does not affect the interference,
while phonon emission from dot 2 suppresses it.
The amount of this suppression, as a function of the bias voltage,
follows the effective e-ph coupling reflecting the phonon density of states and
can be used for phonon spectroscopy.
\end{abstract}

\pacs{71.38.-k, 73.21.La, 73.23.-b}
\maketitle

\section{Introduction}

Detecting dephasing sources in semiconductor quantum-dot devices,
or alternatively investigating  the hallmarks of quantum coherence, is of
much importance for their various applications.
The coherence
of electrons passing through
a quantum dot was demonstrated in a series of experiments,
\cite{yacoby, schuster, kobayashi} in which the dot
has been embedded on an Aharonov-Bohm
\cite{aharonov}(AB) interferometer.
The wave of an electron traversing
the arm of the interferometer  carrying the quantum dot
interferes with the wave passing through the other arm,
resulting in  AB oscillations in the  conductance
as a function of the magnetic flux penetrating the ring.
Dephasing of the interference pattern in  AB interferometers 
has been studied theoretically in several
papers, in conjunction with electronic correlations, 
\cite{akera, konig}
due to coupling with an environmental bath, 
\cite{marquardt,ueda1} or with phonons. \cite{ueda2, ueda3}
Another manifestation of coherence in AB interferometers 
was demonstrated in Ref. \onlinecite{kobayashi}. It was observed that when
high coherence is kept over the whole interferometer,
its conductance shows
an asymmetric shape which is ascribed to a Fano resonance.\cite{fano}
This resonance results from the interference of
tunneling paths through
the continuum of energy levels in the ring
and the leads, with paths passing  through the discrete states in the quantum dot.
Upon increasing the bias voltage \cite{kobayashi} the
Fano resonance gradually takes   the symmetric
Lorentzian shape characterizing a Breit-Wigner resonance of the dot alone.
Interactions of the transport electrons with phonons
(while residing on the dot) have been shown to explain qualitatively the shape change
of the Fano resonance with increasing bias voltages.\cite{ueda2}

Transport electrons passing through a quantum-dot system 
can emit and absorb phonons there.
These electron-phonon (e-ph) interactions can be accompanied by 
energy exchange (inelastic transitions) or not
(i.e., when the same vibration modes are emitted and re-absorbed).
These two types of processes play different
roles in the transport properties of quantum-dot systems.
The latter processes cause the  ``dressing"  of the electrons, resulting in
shifts  of the energy levels in the dot.
For example, these elastic processes narrow the
resonance peak of the linear-response conductance
plotted as a function of the gate voltage (this has
been ascribed in Ref.\ \onlinecite{koch}
to the Franck-Condon blockade, see also Ref. \onlinecite{ora}),
and reduce the height of resonance peak at a finite bias
voltage.
In the case of  inelastic processes, in particular  at a finite bias voltage,
the transport electrons emit or absorb  phonons  and
consequently change their energy states.
The inelastic processes diminish the lifetime of
the electrons on the dot, and hence contribute to the
broadening of the conductance peak.
At zero temperature, and for coupling with optical phonons,
this requires a finite bias voltage.

It is well-known that inelastic processes may lead to 
dephasing. \cite{stern}
Consider for example an AB interferometer carrying a 
quantum dot on one of its arms.
When
the electron emits (or absorbs) a phonon while residing
on the dot, the interference between the waves passing
by the dot and those which do not vanishes, because the
corresponding phonon states are orthogonal to one another.
However, inelastic processes do not always act as a dephasing source.
When the interfering  electron waves interact with the same inelastic scatterer
(emit or absorb the same phonon), that scatterer cannot
be the origin of dephasing. This can be explained \cite{stern}
using the example of  the  h/e and h/2e AB oscillations in the conductance of  
AB  rings. In the  h/e case, one wave goes through one arm of the ring while
the other wave goes through the other.
When the scatterer is located on a certain arm, only one wave interacts with it.
In the h/2e case, one wave goes around the entire ring and
the other wave goes along the reverse  direction.
Therefore, both waves interact with the same scatterer. In this case
the states of the scatterer which are coupled to the waves are the same, and consequently
the scatterer does not harm the interference.

In this paper we examine a different type of 
a quantum-dot interfering device
and show that there again one may encounter 
a situation in which inelastic processes 
do not necessarily destroy   coherence. 
When they do, though, one may exploit 
their effect to extract information on 
the effective coupling with the phonons, 
by measuring the differential conductance 
as a function of the bias voltage.

\begin{figure}
\includegraphics{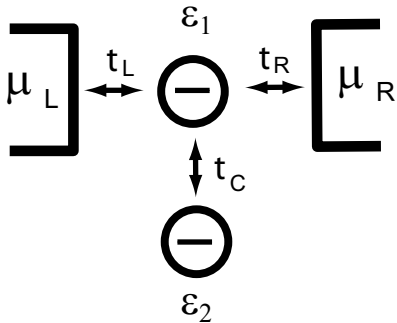}
\caption{\label{fig:fig1}
The T-shaped double-dot system.
The upper dot, dot 1, is connected to the external leads
by the tunneling matrix elements
$t_{\rm L}$ and $t_{\rm R}$; dot 2 is connected, by 
the tunneling matrix element $t_{\rm C}$, solely to dot 1. 
Both dots are represented by a single energy level each,
$\varepsilon_{1}$  and  $\varepsilon_{2}$.
}
\end{figure}

The setup we
consider  is the  T-shaped double-quantum-dot (DQD) structure
depicted schematically in Fig. \ref{fig:fig1}. 
Representing each quantum dot by a single energy level, $\varepsilon_{1}$ and $\varepsilon_{2}$, 
one finds that when electron-phonon interactions are ignored
the  zero-temperature differential conductance  of the device is
\begin{align}
\frac{dI}{dV} =\frac{2e^2}{h} \frac{\alpha \tilde{E}_2^2/4}
{(\tilde{E}_1\tilde{E}_2 - |\tilde{t}_{\rm C}|^2)^2+\tilde{E}^{2}_2/4}\ .
\label{eq:tdot}
\end{align}
Here, 
$\tilde{E}_{i} = (\varepsilon_{i} - eV)/\Gamma$    
[$i=1,2$] 
is the renormalized  on-site energy on each of the 
dots  relative to the bias voltage $eV,$ measured in units of 
the level broadening due to the coupling with the leads, 
$\Gamma=\Gamma _{\rm L} + \Gamma _{\rm R}$.
The partial width resulting  from the coupling with the left (right) 
lead is denoted by $\Gamma_{\rm L}$ ($\Gamma_{\rm R}$).
The asymmetry in these couplings is described by $\alpha=4 
\Gamma _{\rm L} \Gamma _{\rm R}/\Gamma ^2$, and $\tilde{t}_{\rm C} 
= t_{\rm C}/\Gamma$, where $t_{C}$ is the tunneling matrix element 
coupling the two dots (see Fig. \ref{fig:fig1}).
Below, we choose
the chemical potential on the left lead to be  $\mu_{\rm L}
\equiv eV$, and that on the
right one as $\mu_{\rm R}=0$.

\begin{figure}
\begin{center}
\includegraphics[width=6cm]{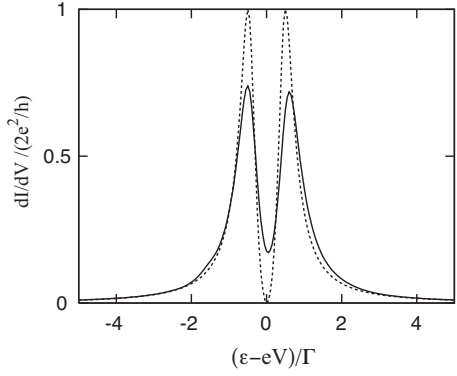}
\caption{
\label{fig:fig2}
The differential conductance $dI/dV$   of a symmetric 
($\varepsilon=\varepsilon_1=\varepsilon_2$, 
$\alpha =1$) DQD structure  as a function of the gate voltage
$\varepsilon$.
In the absence of the e-ph interaction, the conductance 
follows the dotted curve. The solid line is the differential  
conductance when the transport electrons are coupled 
to  acoustic phonons on both dots.
Here  $t_{\rm C} = 0.5 \Gamma$. }
\end{center}
\end{figure}

The dotted line in Fig. \ref{fig:fig2}
is the differential conductance computed from Eq. (\ref{eq:tdot}) for a symmetric DQD,
$\varepsilon = \varepsilon_1 = \varepsilon_2$ and $\alpha =1$. It
has a double-peak structure with
a dip in-between.
The two peaks reflect resonant tunneling  through the bonding and
anti-bonding orbital states of the quantum dots, yielding
$dI/dV=2e^2/h$. At the midpoint between the peaks the 
differential conductance vanishes due to fully destructive interference.
This  structure resembles a Fano resonance in a special case; \cite{fano} indeed the
conductance may be fitted to the Fano line 
shape, $(q+E)^2/(E^2+\gamma^2)$,
 where in our case,  $E =\tilde{E}_{2}$, 
 $\tilde{E}_{1}=0$, $q=0$,
 and $\gamma = 2|\tilde{t}_{\rm C}|^2$.

In the presence of e-ph interactions, this double-peak structure is 
modified significantly, as exemplified by the full curve in Fig.  \ref{fig:fig2}. 
Of particular interest is the effect of those interactions on the destructive 
interference leading to the dip: we find that only the e-ph interaction on dot 2 is 
responsible for the ascent of the dip. Moreover, we argue that
the dependence of this ascent on the bias voltage reflects the effective coupling 
of the e-ph interaction on dot 2, and hence may
serve for phonon spectroscopy. The effect of the e-ph interaction taking place on 
dot 1, together with the one on dot 2, is to decrease the peaks'  height as 
compared to the zero-interaction case.

Below, we study separately the case 
of e-ph interactions with acoustic phonons, and with optical ones.
In semiconductor quantum dots, the electrons interact with the bulk phonons
of the semiconductor base. Then, the electronic interaction with acoustic 
phonons plays a major role at low temperatures since the energies of 
the optical phonons
are much larger ($\sim 36 meV$ for GaAs) than those of the
acoustic phonons. \cite{bruus,keil}
In the case of e-ph interaction with acoustic phonons 
the density of available states is continuous, and indeed  
we find that the dip in the conductance rises up gradually as the bias voltage increases.

The interaction of transport electrons with optical modes 
is considered to be dominant in molecular junctions 
\cite{park, smit, zhitenev, tal, leroy} which {\em do not} 
lie on a substrate. Several theoretical works studying e-ph interactions in
molecular junctions  have employed as a model of the 
molecular bridge a single-level quantum dot coupled to optical vibrations, calculating 
the differential conductance \cite{galperin, mitra, egger, ora, ora2, hod, koch, ryndyk, koch2} 
and the shot noise. \cite{haupt, avriller, schmidt}  
We find that the interaction with optical phonons affects 
the dip once the bias voltage matches the vibrational 
energy on dot 2, but not on dot 1. A delicate issue is the question of the
population of the vibrational states. \cite{ora2} 
When the e-ph interaction is with the acoustic phonons 
of the substrate, one may assume that those are  given 
by the thermal-equilibrium distribution. This assumption 
does not necessarily hold for interactions on molecular bridges, where
the population of the vibrational modes may be determined mainly by the 
transport electrons. Here we will assume that the 
molecular vibrations are thermalized by a 
coupling with the phonon bath of the surrounding.

The organization of the  paper is as follows.
We begin in Sec. II by describing our model and 
presenting  the expression for the differential conductance through the T-shape system. 
The technical details of the calculation are described in the Appendix, 
in particular the treatment of the electron-phonon interaction in the self-consistent Born
approximation.\cite{ueda2,ueda3}
For clarity, we confine ourselves to the case of zero temperature. 
Then, the transport electrons can only emit phonons once the bias 
voltage exceeds the phonon energy.
Section III is devoted to the analysis of our results.
First, we discuss the dephasing-free phonon emission.
We show that the conductance dip is not affected by  phonon emission
from dot 1. Second, we examine the conductance dip 
as a function of the bias voltage $V$.
We show that the amount of decrease in the dip as 
the bias voltage is increased
follows the product of the phonons' density of states 
and the  coupling strength of e-ph interaction. 
We conclude  by a summary/discussion section. 



\section{The model and the calculation method}

\subsection{The model}

Our model system is depicted in Fig. \ref{fig:fig1}, and is described by the Hamiltonian
\begin{align}
H=H_{\rm el}+H_{\rm{e-ph}}+H_{\rm ph}\ .
\end{align}
Omitting the spin indices, the Hamiltonian of the electrons is
\begin{align}
H_{\rm{el}} & =  H_{\rm{L}} + H_{\rm{R}} + H_{\rm{T}} + H_{\rm{D}}.
\end{align}
It consists of the leads'    Hamiltonian
\begin{align}
H_{\rm{L(R)}} & =  \sum_{k } \varepsilon^{}_k
c^{\dagger}_{{\rm{L(R)}}k} c^{}_{{\rm{L(R)}} k},
\end{align}
the DQD Hamiltonian
\begin{align}
H_{\rm{D}} & =  \varepsilon^{}_1 d_{1}^{\dagger} d^{}_{1} + \varepsilon^{}_2 d_{2}^{\dagger} d^{}_{2} ,
\end{align}
and the tunneling Hamiltonian
\begin{align}
H_{\rm{T}} &= \sum_{k} (t^{}_{\rm{L}}
c^{\dagger} _ {{\rm{L}}k} d^{}_1 + {\rm H.c.} )
 + \sum_{k} (t^{}_{\rm{R}} c^{\dagger}_{{\rm{R}}k}
 d^{}_1 + {\rm H.c.} )
  \notag \\
&\quad +t^{}_{\rm C} (d_1^{\dagger} d^{}_2 + \rm{H.c.}).
\label{eq:tunnel}
\end{align}
Here, $c^{\dagger}_{{\rm{L(R)}}k}$ and $c^{}_{{\rm{L(R)}}k}$
denote the creation and annihilation operators of an electron of
momentum $k$ in lead L (R), respectively.
We assume a single energy level $\varepsilon_{1}$ in dot 1 ($\varepsilon_2$ in dot 2),
with the creation and annihilation operators on that level 
being $d_{1}^{\dagger}$, $d^{}_{1}$ ($d_{2}^{\dagger}$, $d^{}_{2}$). 
Electronic correlations are not included in our analysis. The broadening 
of the resonant level on dot 1, $\Gamma=\Gamma_{\rm L}+
\Gamma_{\rm R}$, is given by $\Gamma_{\rm L (R)}= 2 \pi \nu t^2_{\rm L (R)}$, where
$\nu$ is the density of states of the electrons in the leads, 
and $t_{\rm L}$ ($t_{\rm R})$ is the tunneling matrix element connecting the dots to the left (right) lead.

We consider e-ph interactions which are confined to the DQD.  In the 
case of acoustic phonons,  the interaction Hamiltonian reads
\begin{align}
H_{\rm{e-ph}} &= \sum_{\bm{q}} M^{}_{\bm{q},1}
(a^{}_{\bm{q}}+ a^{\dagger}_{-\bm{q}} ) d_{1}^{\dagger}d^{}_{1}
\notag \\
&+  \sum_{\bm{q}} M^{}_{\bm{q},2}
(a^{}_{\bm{q}}+ a^{\dagger}_{-\bm{q}} ) d_{2}^{\dagger}d^{}_{2},
\label{eq:eph}
\end{align}
with the acoustic phonon Hamiltonian being
\begin{align}
H_{\rm ph} =&\sum_{\bm q} \omega^{}_{\bm q} a^{\dagger}_{\bm q}a^{}_{\bm q}.
\end{align}
Here,
$a_{\bm{q}}^{\dagger}$ and $a_{\bm{q}}$ are the creation and annihilation operators of phonons with
momentum $\bm{q}$.
We disregard the possibility of nonlocal e-ph interactions,
$(a_{\bm q} + a^{\dagger}_{- \bm q})d^{\dagger}_{2}d^{}_{1}$,
assuming that the overlap between  the wave functions of dot 1, $|d_{1} \rangle$,
and dot 2, $|d_{2} \rangle$,
is small.
The dispersion relation of the acoustic phonons is taken to be linear,
\begin{align}
\omega_{\bm q} = c_{\rm S}|{\bm q}|
\end{align}
with  the  sound velocity $c_{\rm S}$.
When the e-ph interaction with acoustic phonons originates from the
piezoelectric coupling, the matrix element of the coupling with the electron is \cite{keil, bruus}
\begin{align}
M_{{\bm q}, i} = \lambda_{\bm q} \langle d _i|
e^{i {\bm q}\cdot {\bm r}} | d_i \rangle,
\end{align}
with
\begin{align}
|\lambda_{\bm q}|^2 = g \frac{\pi^2 c_{\rm S}^2 }{ |\bm q|  }.
\end{align}
For example, in GaAs $g = 0.1$.   Because of the oscillating factor
$e^{i {\bm q}\cdot {\bm r}}$, the
e-ph interaction decreases  once the wavelength of the
phonons is smaller than the size of the dot $L_{i}$ ($i=1,2$).  For this reason one may choose
\begin{align}
|M_{{\bm q}, i}|^2 = \frac{\sqrt{2}}{\pi^{1/2}L^2_i}\frac{|\lambda_{\bm q}|^2}{
|{\bm q}|^2 + (1/L_{i})^2}.\label{DOTSIZE}
\end{align}
(The  effect of the mixed product
$M^{}_{{\bm q}, i} M^{\ast}_{{\bm q}, j}$  for $i\neq j$  
is negligible, because  the phase-differnece between 
the two matrix elements that depends on the inter-dot distance diminishes its contribution.)

An ubiquitous  model to describe the interaction with optical phonons
is the
Fr\"{o}hlich Hamiltonian,\cite{mahan} treating the quantum dots as 
Einstein oscillators of frequency $\omega_{i}$, $i=1,2$. In that 
case the electron-phonon Hamiltonian reads
\begin{align}
H_{\rm{e-ph}} &= \zeta^{}_1
(a^{}_{1}+ a^{\dagger}_{1}) d_{1}^{\dagger}d^{}_{1}
+  \zeta^{}_2
(a^{}_{2}+ a^{\dagger}_{2} ) d_{2}^{\dagger}d^{}_{2}
,
\end{align}
and the phonon Hamiltonian is
\begin{align}
H_{\rm ph}& = \omega^{}_{1} a^{\dagger}_{1}a^{}_{1}
+ \omega^{}_{2} a^{\dagger}_{2}a^{}_{2},
\end{align}
where $a_i$ and $a_i^{\dagger}$ are  the creation and
the annihilation operators of phonons on dot $i$ and
$\zeta_i$ is the coupling strength of the optical  e-ph interaction.

Using the Keldysh formalism,\cite{keldysh, caroli, jauho}
the current can be expressed in terms of the Fourier transform of the
retarded Green function of dot 1,
$G^{r}_{11}(t-t^{\prime})= - i\theta(t-t^{\prime})
\langle \{ d^{}_{1}(t),d^{\dagger}_{1}(t^{\prime}) \} \rangle$, \begin{align}
I = \frac{2e}{h}\int d \omega
\Bigl (-\frac{\alpha}{2}\Gamma\Bigr ) [f_{\rm L}(\omega) - f_{\rm R}(\omega)]
{\rm Im}G^{r}_{\rm{11}}(\omega),
\label{eq:cur}
\end{align}
see the Appendix
for details.
Here,  $f_{\rm L(R)}(\omega)=(\exp[\beta (\omega -\mu_{\rm L(R)})]+1)^{-1}$ is
the Fermi distribution function in lead L (R).
In this paper we study the   differential conductance,  given by
\begin{align}
\frac{dI}{dV}=-\frac{2e^2}{h} \frac{\alpha}{2} \Gamma \bigl [  {\rm Im} G^{r}_{11}(eV)
+ \int^{eV}_{0} d\omega {\rm Im} \frac{dG^{r}_{11}(\omega)}{d(eV)}\bigr],
\label{eq:didv}
\end{align}
where the explicit form of $G^{r}_{11}(\omega)$, in  terms of the  self-energies,
is (see the Appendix)
\begin{widetext}
\begin{align}
G^{r}_{11}(\omega) =\frac{\omega- \varepsilon_2}
{(\omega- \varepsilon_1)(\omega- \varepsilon_2) + \frac{i}{2} \Gamma (\omega- \varepsilon_2)
- |t_{\rm C}|^2 - (\omega - \varepsilon_2) \Sigma^{r}_{11}(\omega)
- \frac{|t_{\rm C}|^2}{\omega- \varepsilon_2} \Sigma^{r}_{22}(\omega)}.\label{G11}
\end{align}
\end{widetext}

\section{Results}

\begin{figure}
\begin{center}
\includegraphics[width=6cm]{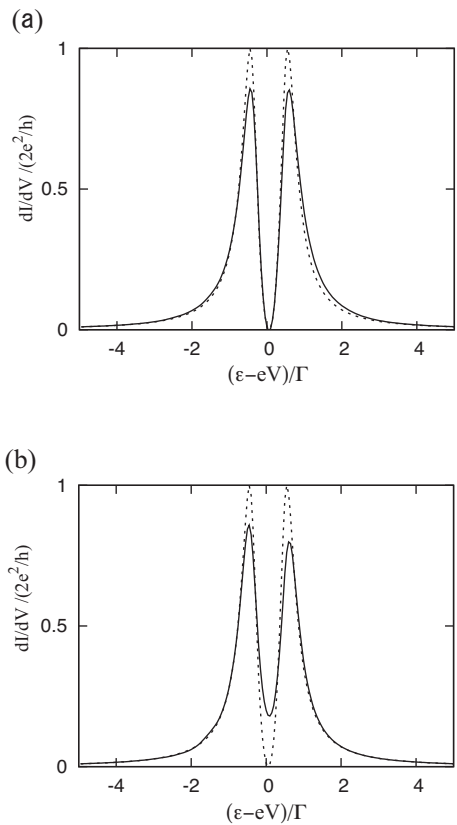}
\caption{\label{fig:fig3}The differential conductance $dI/dV$ (solid lines) as a function of
the gate voltage $\varepsilon$ of  a symmetric DQD
($\varepsilon
=\varepsilon_1=\varepsilon_2$, $\alpha =1$),  for e-ph coupling with acoustic phonons.
The bias voltage is $eV = 2\Gamma$ and
$t_{\rm C} = 0.5 \Gamma$ ($\Gamma$ is the level broadening on dot 1). 
Panel (a): the e-ph interaction is on dot 1 alone; panel (b): the e-ph 
interaction is on dot 2 alone. The dotted curves are the differential 
conductance when the e-ph interactions are absent.
 }
\end{center}
\end{figure}

We begin with  results pertaining to the case in which the electrons are coupled to {\em acoustic} phonons, see Eq. (\ref{eq:eph}).
In Fig. \ref{fig:fig3} we plot the variation of  the differential conductance of a completely symmetric DQD  with   the gate voltage, scaling all energies by $\Gamma$, the level broadening on dot 1.  We present separately results for the case where the  e-ph interaction takes place  on dot 1 alone [panel (a)] or  on dot 2 alone [panel (b)]. Both panels show also the differential conductance obtained in the absence of the coupling with the acoustic phonons. (The parameters chosen are given in the caption  of Fig. \ref{fig:fig3}.)
As is seen from Fig. \ref{fig:fig3}, while phonon emission from dot 1  does not affect the dip in the differential conductance,  it   has  a detrimental effect on $dI/dV$ when it occurs on dot 2. The destructive interference leading to the dip in the differential conductance is severely harmed by the e-ph interaction on dot 2.

To explain this observation, we treat the e-ph interaction to second-order in perturbation theory (note,  however, that the plots in Fig. \ref{fig:fig3} were computed in the self-consistent Born approximation). The dip in $dI/dV$, Eq. (\ref{eq:didv}),  occurs at $\varepsilon = eV$. Then, the first term there vanishes [since at zero temperature ${\rm Im} G^{r}_{11}(eV) = 0$] while the second term yields
\begin{widetext}
\begin{align}
\frac{dI}{dV}
= \frac{2e^2}{h}\frac{\alpha\Gamma}{2} {\rm Im}\Bigl ( \int^{eV}_{0} d\omega
\Bigr [(\omega-eV)^2 + \frac{i}{2} \Gamma (\omega-eV)
- |t_{\rm C}|^2 - (\omega - eV) \Sigma^{r}_{11}(\omega)
- \frac{|t_{\rm C}|^2}{\omega- eV} \Sigma^{r}_{22}(\omega)\Bigr ]^{-2} \nonumber\\
\times\Bigl [ (\omega- eV)^2
\frac{d \Sigma^{r}_{11}(\omega)}{deV} +|t_{\rm C}|^2
\frac{d \Sigma^{r}_{22}(\omega)}{deV} \Bigr]\Bigr ).
\end{align}
Calculating the self-energies appearing in this expression in second-order perturbation theory
we find
\begin{align}
\Sigma^{r(2)}_{11}(\omega)& = \sum_{\bm q} |M_{{\bm q},1}|^2 \int
\frac{d \omega^{\prime}}{2 \pi} |G^{r(0)}_{11}(\omega-\omega^{\prime})|^2\nonumber\\
&\times
\Bigl \{ \frac{\Gamma_{\rm L} f_{\rm L}(\omega-\omega^{\prime})
+ \Gamma_{\rm R} f_{\rm R}(\omega-\omega^{\prime})}
{\omega^{\prime} - \omega_{\bm q} + i0^{+}}
+  \frac{\Gamma_{\rm L} [1-f_{\rm L}(\omega-\omega^{\prime})]
+ \Gamma_{\rm R} [1-f_{\rm R}(\omega-\omega^{\prime})]}
{\omega^{\prime} + \omega_{\bm q} + i0^{+}} \Bigr \},
\end{align}
and
\begin{align}
\Sigma^{r(2)}_{22}(\omega)&=\sum_{\bm q} |M_{{\bm q},2}|^2 \int
\frac{d \omega^{\prime}}{2 \pi} |G^{r(0)}_{12}(\omega-\omega^{\prime})|^2\nonumber\\
&\times
\Bigl \{ \frac{\Gamma_{\rm L} f_{\rm L}(\omega-\omega^{\prime})
+ \Gamma_{\rm R} f_{\rm R}(\omega-\omega^{\prime})}{\omega^{\prime}
 - \omega_{\bm  q} + i0^{+}}
+  \frac{\Gamma_{\rm L} [1-f_{\rm L}(\omega-\omega^{\prime})]
+ \Gamma_{\rm R} [1-f_{\rm R}(\omega-\omega^{\prime})]}{\omega^{\prime}
+ \omega_{\bm q} + i0^{+}}\Bigr  \} .
\end{align}
A straightforward calculation shows that
 $d\Sigma^{r(2)}_{11}(\omega)/d(eV) $  at $\epsilon =eV$ vanishes [this follows from Eq. (\ref{G11})]. Hence, when the e-ph interaction on dot 2 vanishes, so does the differential conductance at the mid-point. On the other hand,
\begin{align}
 \frac{\Sigma^{r(2)}_{22}(\omega)}{deV} = &
-\sum_{\bm q} \frac{|M_{{\bm q},2}|^2}{2 \pi}
\frac{\Gamma_{\rm L}}{|t_{\rm C}|^2}
\Bigl ( \frac{1}{\omega-\omega_{\bm q}
-eV + i0^{+}}
-\frac{1} {\omega+\omega_{\bm q}- eV + i0^{+}}\Bigr)\nonumber\\
&
=\frac{g}{2(2 \pi)^{1/2}}
\frac{\Gamma_{\rm L}}{|t_{\rm C}|^2}\bigl (\frac{c_{\rm S}}{L_{2}} \bigr )^2
\frac{\omega-eV}{(\omega-eV)^2 + (c_{\rm S}/L_{2})^2}
\label{eq:phout},
\end{align}
\end{widetext}
yielding that  when phonon emission takes place on dot 2 the dip in the differential conductance is modified.

\begin{figure}
\begin{center}
\includegraphics[width=6cm]{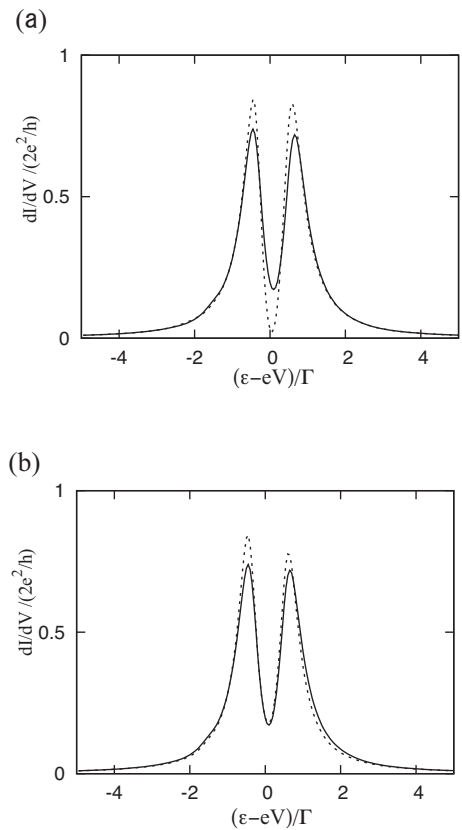}
\caption{\label{fig:fig4} The differential conductance $dI/dV$ as a function of
the gate voltage $\varepsilon$ of  a symmetric DQD
($\varepsilon
=\varepsilon_1=\varepsilon_2$, $\alpha =1$),  for e-ph coupling with acoustic phonons. 
The bias voltage is $eV = 2 \Gamma$  and
$t_{\rm C} = 0.5 \Gamma$ ($\Gamma$ is the level broadening on dot 1). 
Panel (a): changing the size of dot 2. The solid line is the differential 
conductance when both dot sizes are equal,
$L_{1}= L_{2}=c_{\rm S}/{(2.0 \Gamma})$ and the dotted line is for the 
case $L_{1}= c_{\rm S}/{(2.0 \Gamma})$ and  $L_{2}=c_{\rm S}/{(0.5 \Gamma})$. Panel
(b): changing the size of dot 1. The solid line is the differential 
conductance when both dot sizes are equal,
$L_{1}= L_{2}=c_{\rm S}/({2.0 \Gamma})$ and the dotted line is f
or the case $L_{1}= c_{\rm S}/{(0.5 \Gamma})$ and  $L_{2}=c_{\rm S}/{(2.0 \Gamma})$.
}
\end{center}
\end{figure}

\begin{figure}
\begin{center}
\includegraphics[width=6cm]{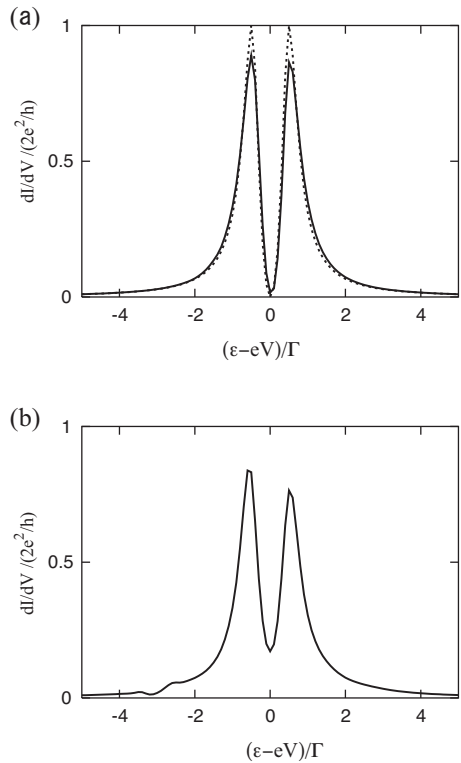}
\caption{\label{fig:fig5}  The differential conductance $dI/dV$ as a function of
the gate voltage $\varepsilon$ of  a symmetric DQD
($\varepsilon
=\varepsilon_1=\varepsilon_2$, $\alpha =1$),  for e-ph coupling with optical phonons. Here
$t_{\rm C} = 0.5 \Gamma$ ($\Gamma$ is the level broadening on dot 1), and  the
strengths of e-ph coupling on the dots are $\zeta_1=\zeta_2 = 0.3 \Gamma$.
$\delta$ introduced in Eqs. (\ref{eq:delta1}) and (\ref{eq:delta2})  is
$\delta = 0.5 \Gamma$.
Panel (a):  the phonon energy in dot 1 (dot 2) is $\omega_1 = \Gamma$
($\omega_2 = 3\Gamma$), and the
bias voltage is  $eV= \Gamma$.
The dotted line indicates the conductance in the absence of e-ph
interaction.
Panel (b): the same as in (a) with   $eV=3\Gamma$.
}
\end{center}
\end{figure}

Since the e-ph interaction on dot 1 does not harm the interference around $\epsilon =eV$, 
one may say that it does not cause dephasing.  One may monitor this 
(zero-temperature) dephasing-free phonon emission by varying the size of 
the DQD. As is mentioned above [see Eq. (\ref{DOTSIZE})], the
efficacy of the e-ph interaction with acoustic phonons depends on the dot size:  
it  decreases as the dot size increases. The effect of  the two dots' sizes on 
the differential conductance is investigated in Fig.
\ref{fig:fig4}. In both panels, the solid curves pertain to the case of equal-size 
dots. The dotted curve in panel (a) shows the modification in the differential 
conductance brought about by {\em increasing} the size of dot 2 (thus making 
the phonon emission there less efficient). It is clearly seen that the decrease in  
$dI/dV$ towards the dip is severely disturbed. On the other hand,  when the 
size of dot 1 is increased [panel (b)] the differential conductance is almost unchanged.

Next we consider the differential conductance in the case 
where the electrons are coupled to {\em optical} phonons. 
The main difference
between this interaction and the coupling with acoustic phonons 
discussed above, is that now the transport electrons can emit  
real phonons and change their energies only when the bias 
voltage exceeds the phonon frequency (at zero temperature).  
This is portrayed in
 Fig.\ \ref{fig:fig5}.  Switching-on the coupling to optical phonons 
 when the vibration frequency on dot 2 is {\em larger} than the bias 
 voltage almost makes no difference in the shape of the differential 
 conductance [see panel (a) of Fig. \ref{fig:fig5}]. The modifications 
 in the peaks' heights are due to elastic processes taking place on both dots. 
 On the other hand, when the bias voltage is large enough (as compared to 
 the vibration frequency on dot 2),  there is
a significant effect on the dip in $dI/dV$, see panel (b) of Fig. \ref{fig:fig5}.

\begin{figure}
\begin{center}
\includegraphics[width=6cm]{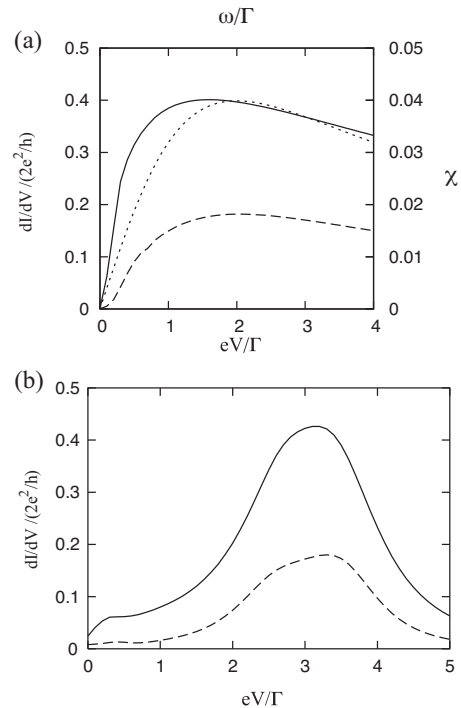}
\caption{\label{fig:fig6}The differential conductance $dI/dV$
as a function of the bias voltage $V$
 for a symmetric DQD ($\varepsilon
=\varepsilon_1=\varepsilon_2$, $\alpha=1$),  at the conductance dip
($\varepsilon-eV=0$).
The tunnel coupling between the dots is $t_{\rm C}=0.3 \Gamma$ (solid lines)
and $t_{\rm C}= 0.5 \Gamma$ (dashed lines). Panel (a): coupling with acoustic 
phonons; the dotted line is
$\chi(\omega)$ [Eq. (\ref{CHI})] as a function of $\omega$ (marked on the right-side 
vertical axis). Panel (b): coupling with optical phonons. The vibration energy in dot 1  
is $\omega_1 = \Gamma$ and on dot 2 is
$\omega_2 = 3\Gamma$. }
\end{center}
\end{figure}

 Finally we study the possibility to use the  dephasing effect of the e-ph interactions 
 on dot 2 for phonon spectroscopy. Figure \ref{fig:fig6} depicts the dependence of 
 the conductance at the dip on the bias voltage for coupling with acoustic 
 phonons [panel (a)] and optical phonons [panel (b)]. The coupling of the 
 electrons to acoustic phonons is characterized by the spectral function 
 obtained from the product of their density of states and the (acoustic) e-ph matrix element squared,
 \begin{align}
\chi(\omega) & =  \sum_{\bm q}|M_{\bm q,2}|^2 \delta(\omega - \omega_{\bm q}) \\
& = \frac{g}{(2 \pi)^{1/2}}\bigl (\frac{c_{\rm S}}{L_{2}} \bigr )^2
\frac{\omega}{\omega^2 +(c_{\rm S}/L_{2})^2}.\label{CHI}
\end{align}
Only the spectral function on dot 2  is presented, since 
(as was elaborated upon above) the conductance dip is 
affected by the e-ph interaction on that dot. The function 
$\chi (\omega )$ is shown by the dotted curve of Fig. \ref{fig:fig6}. 
Although it does not coincide with  the conductance 
curve it does follow it, in particular at higher values of the voltage.

Panel (b) of Fig. \ref{fig:fig6} shows the value of the conductance 
dip (as a function of the bias voltage) in the case where the electrons 
are coupled to optical phonons. The prominent feature here is the peak 
obtained when the bias voltage matches the vibration 
frequency on dot 2. The decrease in the 
conductance dip becomes more pronounced as the tunnel coupling between 
the two dots, $t_{\rm C}$, is {\em decreased}. This is because when this coupling is strong, separate effects of 
the dots on the conductance is more blurred. On the other hand, a smaller
value of this coupling enables the manifestation of the e-ph interaction on dot 2 
to become more distinguished. We do not plot the spectral function corresponding
to the case of optical phonons since our calculation does not take into consideration
the origin of the life time of these phonons. The broadening $\delta$  introduced
in Eqs. (\ref{eq:delta1}) and  (\ref{eq:delta2}) is a free parameter, that in fact
can be extracted by fitting the curves in Fig. \ref{fig:fig6} (b) to the experimental data.

\section{Discussion}

In summary, we
have studied the effect of electron-phonon interactions
in a T-shaped DQD, employing the self-consistent Born approximation.
The differential conductance of this device is characterized by 
the appearance of a double-peak structure due to interference. 
The dip in-between the two peaks is quite sensitive to the dephasing effect of
phonon emission.  We find that while phonon emission from  dot 1 does not affect the
conductance dip,  phonon emission from dot 2 leads to its ascent.  
Therefore, the electron-phonon interaction on dot 2, in particular the 
effective e-ph coupling given by the  product of the phonons' density 
of states and the e-ph interaction, can be probed electrically,  by 
monitoring the variation of the conductance dip.

The conductance dip  always appears at $\varepsilon_2 = eV$,  
even when the DQD is asymmetric (for example, $\varepsilon_1 \neq \varepsilon_2$). 
Since the behavior of
the self energy, which is responsible for the appearance of the dip,
remains the same also in this case,
our conclusions should hold for the asymmetric double-quantum-dot system.
On the other hand, we expect that a finite temperature will smear the conductance dip (even in the absence of the
electron-phonon interactions). When the temperature differs from zero,
the behavior of the self energy is slightly (at low temperature) modified,
and the distinction between the effects of the e-ph interactions on dot 1
and those on dot 2 is less clear. A finite temperature 
will also facilitate phonon absorption processes. 
However, these alone are not expected to modify 
our main results considerably.
One may hope though, that phonon spectroscopy 
via the monitoring of the conductance dip is 
still possible at low enough temperatures.
This in particular is so since we propose to monitor the differential 
conductance at finite bias voltages. 

An interesting point is 
the experimental possibility to have the electron-phonon 
interaction taking place mainly on one of the dots forming the DQD system.
A phonon bath may be realized by an electronically insulating hard substrate.
It is harder to imagine how a quantum dot can be effectively detached 
from a phonon bath and become ``floating"; this seems to be more realistic in the case of a  molecular junction, for which, under these circumstances, the relevant vibrational modes are optical. Then, a sufficiently high phonon frequency (higher than the bias voltage required for the dip) will satisfy the above requirement.

\begin{acknowledgments}
This work was partly supported by the Strategic Information
and Communications R\&D Promotion Program (SCOPE) from the
Ministry of Internal Affairs and Communications of Japan,
by a grant-in-aid for scientific research from
the Japan Society for the Promotion of Science,  by the German Federal Ministry of
Education and Research (BMBF) within the framework of the
German-Israeli project cooperation (DIP), and by the US-Israel
Binational Science Foundation (BSF).
\end{acknowledgments}

\appendix

\section{Details of the calculation}

\label{DETAILS}

Our calculation requires all four Keldysh Green
functions \cite{keldysh, caroli, jauho}
belonging to the DQD:   the time-ordered one
\begin{align}
G^{t}_{ij}(t-t^{\prime})
&= - i\langle \mathcal{T} d^{}_i(t) d^{\dagger}_j
(t^{\prime})  \rangle,
\end{align}
the anti time-ordered one,
\begin{align}
G^{\tilde{t} }_{ij}(t-t^{\prime})
&= - i\langle \tilde{ \mathcal{T}} d^{}_i(t) d_j^{\dagger}
(t^{\prime})  \rangle,
\end{align}
with $\mathcal{T}$ ($\tilde{\mathcal{T}}$) being the
time-ordering (anti time-ordering) operator,
and the lesser and greater Green functions,
\begin{align}
G^{<}_{ij}(t-t^{\prime})& = i \langle
d_j^{\dagger}(t^{\prime}) d_i(t) \rangle,
\nonumber\\
G^{>}_{ij}(t-t^{\prime})& =
- i \langle d_i(t)d_j^{\dagger}(t^{\prime}) \rangle .\label{4G}
\end{align}
Here $i$ and $j$ take the values 1 and 2.

It is convenient to present the Fourier transforms of  these Green functions
 in a matrix form
\begin{align}
\bm{G}^{\gamma}(\omega) = \begin{pmatrix} G^{\gamma}_{11}(\omega) &
G^{\gamma}_{12}(\omega) \\ G^{\gamma} _{21}(\omega)&
G^{\gamma}_{22}(\omega) \end{pmatrix}
\label{eq:matrix}
\end{align}
where $\gamma=t$, $\tilde{t}$, $<$, or $>$.  
The retarded and advanced Green functions follow from these functions,
 \begin{align}
\bm{G}^{r}(\omega) =
\bm{G}^{t}(\omega) - \bm{G}^{<}(\omega).
\label{eq:grtl}
\end{align}
All the necessary  Green functions are obtained
from the corresponding Dyson equations, in which there
appears  the self-energy due
to the e-ph interaction,  $\Sigma$,
\begin{align}
\bm{G}^{r}(\omega) &=  \bm{G}^{r(0)}+\bm{G}^{r(0)}(\omega) \bm{\Sigma}^r(\omega)
\bm{G}^r(\omega),\label{eq:rDyson}
\end{align}
and the lesser Green function of the DQD
\begin{align}
&\bm{G}^{<}(\omega) = \bm{G}^{r}(\omega)
\bm{\Sigma}^{<}(\omega) \bm{G}^{a}(\omega)\nonumber\\
&+[1 + \bm{G}^{r}(\omega) \bm{\Sigma}^{r}(\omega)]
\bm{G}^{<(0)}(\omega)
[1 + \bm{\Sigma}^{a}(\omega)
\bm{G}^{a}(\omega) ] ,
\label{eq:lDyson}
\end{align}
where  $\bm{G}^{(0)}$ is the Green function
in the absence of the coupling to the phonons.

We treat the e-ph interactions in the self-consistent
Born approximation. \cite{ueda2, ueda3} Since we 
focus on the effect of the inelastic electron-phonon processes, we discard the Hartree term.
When the transport electrons are coupled to  acoustic phonons,
the required  self-energies are
\begin{align}
\Sigma^{t}_{ii} = \frac{i}{2 \pi}
\sum_{\bm{q}} |M_{\bm{q},i}|^2
\int d \omega^{\prime} G_{ii}^{t}(\omega-\omega^{\prime})
D^{t}(\bm{q}, \omega^{\prime}),\label{eq:self1}
\end{align}
and
\begin{align}
\Sigma^{<}_{ii} = \frac{i}{2 \pi}
\sum_{\bm{q}}|M_{\bm{q},i}|^2
\int d \omega^{\prime} G_{ii}^{<}(\omega-\omega^{\prime})
D^{<}(\bm{q}, \omega^{\prime}).
\label{eq:self2}
\end{align}
Here $D$  denotes the Fourier transform of the zero-order 
(i.e., in the absence of the coupling to the electrons) phonon Green functions,
\begin{align}
D^{t}({\bm{q}}, \omega) &= - 2 \pi i
[N_{\bm{q}} \delta (\omega + \omega_{\bm{q}})
+ N_{\bm{q}} \delta(\omega - \omega_{\bm{q}})] \notag \\
&\quad  + \frac{1}{\omega - \omega_{\bm{q}} + i 0^{+} }
- \frac{1}{\omega + \omega_{\bm{q}} - i 0^{+} },
\end{align}
and
\begin{align}
D^{<}({\bm{q}}, \omega) &=
-2 \pi i [(N_{\bm{q}} + 1) \delta(\omega + \omega_{\bm{q}}) \notag \\
&+ N_{\bm{q}} \delta(\omega - \omega_{\bm{q}})],
\end{align}
where $N_{\bm q}$ is the phonon population of mode ${\bm q}$. 
It is implicitly assumed that the vibrational modes are equilibrated 
by the coupling to another heat bath, such that their population is given by
Bose-Einstein distribution function,
$1/[\exp (\beta \omega_{\bm{q}})-1]$. Then,
$N_{\rm q}=0$ at zero temperature. The self-energies
 $\Sigma^{r}_{ii}(\omega)$ and $\Sigma^{<}_{ii}$ are determined
by solving Eqs.\ \eqref{eq:self1} and \eqref{eq:self2}
self-consistently.

When the electrons are coupled to optical phonons,
the self-energies are
\begin{align}
\Sigma^{t}_{ii} = \frac{i}{2 \pi}
\zeta_{i}^2
\int d \omega^{\prime} G_{ii}^{t}(\omega-\omega^{\prime})
D^{t}_{i}(\bm{q}, \omega^{\prime}),\label{eq:self1-opt}
\end{align}
and
\begin{align}
\Sigma^{<}_{ii} = \frac{i}{2 \pi}
\zeta_{i}^2
\int d \omega^{\prime} G_{ii}^{<}(\omega-\omega^{\prime})
D^{<}_{i}(\bm{q}, \omega^{\prime}),
\label{eq:self2-opt}
\end{align}
where the Fourier transforms of phonon Green functions
at zero temperature are
\begin{align}
D^{t}_{i}(\omega)=\frac{1}{\omega - \omega_{i} +  i \delta}
- \frac{1}{\omega +\omega_{i} - i \delta},
\label{eq:delta1}
\end{align}
and
\begin{align}
D^{<}_{i}(\omega)=\frac{1}{\omega +\omega_{i} +  i \delta}
- \frac{1}{\omega +\omega_{i} - i \delta}.
\label{eq:delta2}
\end{align}
Here $\delta$ is relaxation rate of the
Einstein phonon mode due to
the coupling with the  surrounding  bulk phonons.
As in the acoustic-phonon case,  $\Sigma^{r}_{ii}(\omega)$ and $\Sigma^{<}_{ii}$
are determined by solving
Eqs. (\ref{eq:self1-opt}) and (\ref{eq:self2-opt}) self-consistently.

The Green functions of the DQD, in particular
$G^{r}_{11}$, determine the expression for the current.
The operator of the current between the left  lead and   the quantum dots
is given
by the time derivative of the electron number operator in that lead,
$N_{\rm{L}} = \sum_{k} c^{\dagger}_{{\rm L} k}c^{}_{{\rm L} k}$,
\begin{align}
I_{\rm{L}} = - 2e \langle {\dot{N}_{\rm{L}}} \rangle
&= -2ie \langle [H, N_{\rm{L}}]\rangle =  4 e {\rm Re}
\sum_{k} t_{\rm{L}}
G^{<}_{{\rm 1, L} k}(t,t)\nonumber\\
& =  \frac{4e}{h}{\rm Re} \int d \omega
t_{\rm{L}} \sum_k
G^{<}_{{\rm 1, L} k}(\omega) \ ,
\end{align}
adding a factor of 2 for the spin components.
Here
$G^{<}_{{\rm 1, L}k}(t-t^{\prime})
= i \langle c^{\dagger}_{{\rm L} k}(t^{\prime}) d^{}_{1}(t) \rangle$
is the lesser Green function
and
$G^{<}_{{\rm 1, L}k}(\omega)$ is its Fourier transform.
The current from lead R to the quantum dots, $I_{\rm R}$,
is obtained in the same way.

Using the equation-of-motion method  \cite{jauho}
the current from lead L (R) is rewritten as
\begin{align}
I_{\rm L(R)} =&
\frac{4e}{h}\int d \omega \biggl[ -\frac{\Gamma_{\rm L(R)}}{2}
{\rm Im}G^{<}_{\rm{11}}(\omega)\nonumber\\
&
- \Gamma_{\rm  L(R) } f_{\rm L}(\omega) {\rm Im }
G^{r}_{\rm{11}}(\omega) \biggr],
\end{align}
where $G^{r}$ denotes the retarded Green function.
The net current through the DQD system  is hence
\begin{align}
I &=\frac{1}{2}\Bigl (
I_{\rm L}-I_{\rm R}\Bigr )
\nonumber\\
&=\frac{4e}{h}\int d \omega
 \biggl \{ -\frac{1}{4}\Bigl ( \Gamma_{\rm L}-\Gamma_{\rm R} \Bigr )
{\rm Im}G^{<}_{\rm{11}}(\omega)
\nonumber\\
&-\frac{1}{2}\Bigl (\Gamma_{\rm  L} f_{\rm L}(\omega)-
\Gamma_{\rm  R} f_{\rm R}(\omega)
 \Bigr )
 {\rm Im }
G^{r}_{\rm{11}}(\omega) \biggr\}. \label{eq:i}
\end{align}
Using the relation $I_{\rm L} + I_{\rm R} = 0$  which
follows by charge conservation,   and assuming that the widths $\Gamma_{\rm L}$ and
$\Gamma_{\rm R}$ do not vary significantly with energy,
we can express the lesser Green function in terms of the retarded one,
\begin{align}
{\rm Im} G^{<}_{11}(\omega) = -\frac{2[\Gamma_{\rm L} f_{\rm L}(\omega) +
\Gamma_{\rm R} f_{\rm R}(\omega)]}{\Gamma_{\rm L} + \Gamma_{\rm R}}
 {\rm Im }
 G^{r}_{\rm{11}}(\omega). \label{eq:cc}
\end{align}
Inserting Eq.\ \eqref{eq:cc} in Eq. \eqref{eq:i} leads  to Eq. \eqref{eq:cur}.

\end{document}